\documentclass[%
 reprint,
 amsmath,amssymb,
 aps,
 nofootinbib
]{revtex4-2}
\makeatletter

\renewcommand*{\p@subsection}{}

\renewcommand*{\p@subsubsection}{}
\makeatother
\usepackage{float}
\usepackage{graphicx}
\usepackage{dcolumn}
\usepackage{bm}
\usepackage{soul} 
\usepackage{color}
\definecolor{vs}{rgb}{0.1,0.4,0.1}                  

\newcommand{\del}[1]{}                             
\usepackage{hyperref}
\usepackage{verbatim}
\newcommand\wordcount{\verbatiminput{\jobname.sum}}
\begin{document}
\begin{minipage}[h]{\textwidth}
  Laser Photonics Rev. 2300512 (2023). \\
  \protect
  \url{https://doi.org/10.1002/lpor.202300512}
  \mbox{}\\
\end{minipage}
\title{The Poynting vector field singularities: Effects of symmetry and its violation}


\author{Michael I. Tribelsky}
\email[
]
{\mbox{E-mail: mitribel\_at\_gmail.com (replace ``\_at\_" by @)}
}
\homepage[]{https://polly.phys.msu.ru/en/labs/Tribelsky/}
\affiliation{
M. V. Lomonosov Moscow State University, Moscow, 119991, Russia}
\affiliation{Center for Photonics and 2D Materials, Moscow Institute of Physics and Technology, Dolgoprudny 141700, Russia}

\date{\today}

\begin{abstract}
{The phenomenological theory revealing the generic effects of the problem symmetry, its violation, and energy conservation law on the singularities of the Poynting vector field is presented. The bifurcation scenario of their formation (annihilation) under variations of the problem parameters is elucidated. The results describe the singularities in scattering a linearly polarized plane electromagnetic wave. However, they are valid for any configuration of the incident beam at its scattering by a subwavelength particle. The author shows that topological changes in the pattern of the Poynting vector field occur through a finite number of pitchfork bifurcations. It means that the patterns are topologically stable under variations of the problem parameter(s) that lie between the bifurcation values. The latter ensures that the discussed topological properties of the problem are robust to weak symmetry violation, which is inevitable in any actual experiment. The general consideration is illustrated by a detailed study of singularities in scattering by an infinite right circular germanium cylinder. The results open the possibility of fitting and controlling radiation patterns on subwavelength scales important for various nanotechnologies.
}

\end{abstract}


\maketitle
%
%
\mbox{}
\mbox{}\\
\mbox{}\\
Accepted: 2023-12-07; Received: 2023-06-05;\\
Revised: 2023-10-17;
\section{Introduction\label{sec:Intr}}

The fundamental role of singularities in determining the topological properties of any field is well-known. For this reason, they attract the close attention of researchers. In the case of electromagnetic fields, remarkable achievements in the study of singularities and their applications are presented in numerous publications; see, e.g.,~\cite{Wang2004,Bashevoy2005,Mokhov2006,Tribelsky2006,Lukyanchuk2007,Mokhun2007,Dennis2009,Novitsky2009,Yue2019,Angelsky2021} and reference therein. 

The electromagnetic field singularities may be related to those in the fields $\mathbf{E}$ (electric), $\mathbf{H}$ (magnetic), and $\mathbf{S}$ (the Poynting vector). Though all these fields are interconnected, each has its own peculiarities, and the corresponding singularities should be discussed individually.

{The topology of the Poynting vector field is not only of purely academic interest but also essential for many applications. The point is that this field and its divergence describe electromagnetic energy flow and energy release due to dissipation, respectively. Knowledge of the details of these processes and the ability to control and tailor the field pattern on subwavelength scales opens the door to breakthroughs in various nanotechnologies. Specifically, in biology and medicine, it enables nanosurgery~\cite{Nanosurgery2005} and opens new perspectives for studying chromosomes~\cite{berns2020laser}. Nanostructuring and nanomelting by laser pulses can produce metasurfaces with desired properties~\cite{nanomelting2011}; precise control of electromagnetic energy density at subwavelength scales is fundamental in developing nanoscale optoelectronic devices~\cite{Gaweda2009}. Singularities of the Poynting vector field affect the angular momentum application point. It can be used to develop a new type of optical traps~\cite{Mokhun2008}. A tailored field pattern in the subwavelength range can considerably improve the accuracy and sensitivity of the study of individual nanoparticles and molecules~\cite{nie1997probing}. Spatially inhomogeneous field distribution is required for various versions of optical tweezers~\cite{fan2014light}, nanoradars~\cite{Lapshina2012}, etc. The list can be easily extended; see, e.g., Handbook~\cite{Bhushan2017}} 

{On the other hand, direct experimental measurements of the electromagnetic field at subwavelength scales are challenging~\cite{Kapitanova2017}, especially if one has to measure the field inside a solid body. In this case, information obtained from a theoretical inspection of the singularities may help significantly.}


{The Poynting vector field lines and their singularities are the subjects of many publications, starting from the pioneer paper by Bohren~\cite{Bohren1983}; see, e.g., Refs.~\cite{Wang2004,Bashevoy2005,Tribelsky2006,Lukyanchuk2007,Lukyanchuk2007a,Canos2021,Tribelsky2022a,Yue2022,Geints2022,Geints2022a}. In particular, work~\cite{Novitsky2009} presents a detailed classification of the Poynting vector field singularities.}

{However, most studies relate to examining features of complex configuration fields. Many interesting, significant, and often unexpected results have been obtained. They include the singularities of nonparaxial Bessel beams~\cite{Gao2014}; singularities in superoscillating fields~ \cite{Berry2019}; singularities due to interference with toroidal modes~\cite{ospanova2021}; the appearance of a reverse (relative to the incident wave) energy flow near the axis of the sharply focused vortex beam~\cite{Kotlyar2019}, etc. As a rule, the theoretical description of these features implies cumbersome computations involving such concepts as topological charges~\cite{Berry2004}, geometric phases of Pancharatnam–Berry~\cite{berry1984,Cohen2019}, etc.}

Meanwhile, the fields $\mathbf{E}$, $\mathbf{H}$, and $\mathbf{S}$ satisfy Maxwell's equations and the energy conservation law, respectively. Though everybody knows it, the fact that these conditions impose {certain} constraints on the features of the singular points has {not} been either noticed or given sufficient attention. As a result, even in the most straightforward cases of scattering of a {linearly polarized plane} electromagnetic wave, some important generic properties of the singularities have not yet been revealed. 

{In our previous publications, we discuss some of them related to the effects of the spatial dimension~\cite{Tribelsky2022} and dissipative properties of the scatterer~\cite{Tribelsky2022_Nanomat_diss} on the features of a single, separated singularity. The consideration is restricted by the highly symmetric light scattering problems by a sphere and infinite right circular cylinder. For the latter, only pure TE and TM polarizations have been discussed.} 

{In the meantime, the questions are: ``How do these singularities come into being (disappear) under variations of the problem parameters?'' and ``What laws govern and control these phenomena?'' To answer them, one should uncover the bifurcation scenarios. Despite the apparent importance of these questions, the corresponding study has never been done.}    

{Moreover, in actual experimental cases, symmetry is always violated. Then, the questions: ``How does symmetry violation affect the properties of the Poynting vector field singularities?'' and ``How robust is the topological structure of the singularities to symmetry violation?'' arise.} 

{In the present paper, we try to answer all these questions. We show that the symmetry violation produces qualitatively new effects: The field line pattern becomes essentially three-dimensional, and most singular points turn regular. In these cases, entities, which we call {\it false singular points\/} may come into being. These points are singular in two-dimensional projections of the field lines but regular in the three-dimensional space. {Similar ``false'' singularities are often encountered in various optical problems, for example, in the theory of paraxial beams~\cite{bekshaev2007,bekshaev2011}. The difference between the mentioned ``conventional'' false singularities and the ones discussed in the present paper is that while for the former, the direction of the non-zero Poynting vector component is determined by the beam symmetry, in our case, the scatterer symmetry plays this role; see below. We classify the types of bifurcations related to the creation (annihilation) of singular points (both true and false) and develop a phenomenological theory of the bifurcation phenomena.}

We illustrate the general discussion with a detailed study of light scattering by an infinite right circular cylinder. We consider the normal incidence and an arbitrary orientation of the incident wave polarization plane against the cylinder axis. {This problem formulation is a simple and analytically tractable example of the generic case of a system with several symmetry groups when a part of them is violated in a controllable manner while the others remain unperturbed. Specifically, variations of the angle $\alpha$ between the vector $\mathbf{E}$ of the incident wave and the cylinder axis make it possible to observe a smooth violation of the mirror symmetry against the cylinder's normal cross section plane. At $\alpha = 0$ and  $\alpha = 90^\circ$ we have the highly symmetric pure TM or TE polarizations, respectively. Departures of $\alpha$ from these two extreme values give rise to a mixed TM--TE case with violated mirror symmetry. The violation range smoothly increases with an increase in the departure, achieving the maximal value at  $\alpha=45^\circ$. In contrast, translation symmetry along the cylinder axis remains valid at any value of $\alpha$.} 

To verify the universality of the obtained bifurcation features, we also inspect bifurcations initiated by the change of the cylinder radius for the pure TM polarization of the incident wave. Comparing the two cases (the fixed radius and variable orientation of the polarization plane vs. pure TM polarization and variable radius) confirms the results' generic nature.

{We perform the calculations using the corresponding exact problem solutions in all specific examples illustrating the general discussion. This approach makes it possible to readily keep the accuracy of the calculations sufficient to resolve the finest details of the phenomena under consideration.} 

{Finally, we must emphasize an essential feature of the singularities discussed in the present work that distinguishes them from most publications in the field. While the developed phenomenological approach is quite general and can be applied to any singularity, the main challenge for the above-mentioned nanotechnological applications is to generate a field pattern in the subwavelength range. In conventional wave optics, the minimum spatial scale for a field pattern is the wavelength order of magnitude. Thus, wave optics cannot help to accomplish the task. On the other hand, when a subwavelength particle scatters light, the particle's size determines the characteristic scale of the field pattern in the particle and its vicinity. In this region, we can always consider the incident wave of any configuration to be locally plane and the singularities to arise spontaneously since a plane wave is spatially homogeneous and cannot enforce any singularities. Therefore, singularities at a subwavelength particle light scattering are a convenient (and, perhaps, the only) tool to tailor and control field patterns with subwavelength spatial scales. Because of that, the light scattering by a subwavelength cylinder discussed in this paper is more significant than just an example illustrating the phenomenology.}

The paper has the following structure. In Section~\ref{sec:formulation}, we give the problem formulation and introduce our method to attack it. In Section~\ref{sec:symmetry}, we reveal the symmetry of the fields $\mathbf{E}$, $\mathbf{H}$, and $\mathbf{S}$ at light scattering by a cylinder, under the {conditions} specified above. In Section~\ref{sec:steram}, we discuss the general features of the Poynting vector field lines (streamlines) and singularities. Section~\ref{sec:phenomenology} is the key one: there, we develop a phenomenological theory of the Poynting vector field singularities and inspect how the symmetry and energy conservation law affect them. In Section~\ref{sec:examples}, we present specific examples of light scattering by a germanium cylinder, illustrating the developed phenomenology. In Section~\ref{sec:conclus}, we briefly formulate the main results of the study.

\section{Problem formulation and Method\label{sec:formulation}}

\begin{figure}
  \centering
  \includegraphics[width=.5\columnwidth]{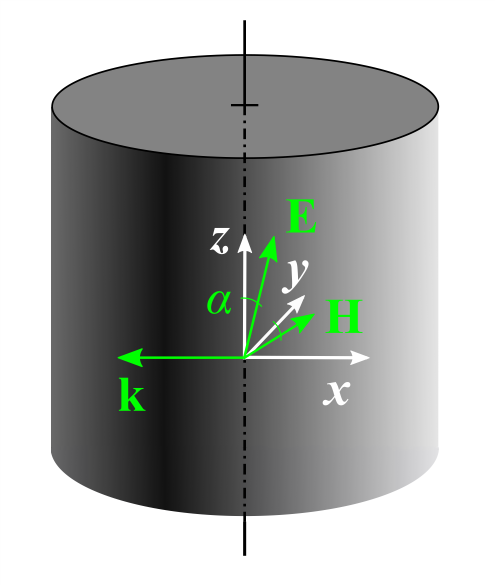}
  \caption{Mutual orientations of the cylinder, the coordinate system, and {the} vectors $\mathbf{k}$, $\mathbf{E}$, and $\mathbf{H}$ of the incident wave; $\alpha$ is the angle between the $z$-axis and vector $\mathbf{E}$ equal to the one between the $y$-axis and vector $\mathbf{H}$.}\label{fig:Orientation}
\end{figure}
We consider an infinite right circular cylinder with a complex permittivity $\varepsilon=\varepsilon'+i\varepsilon''$ and permeability $\mu = 1$ (which is common in the optical frequency range) in a vacuum, irradiated by a linearly polarized plane monochromatic electromagnetic wave. The temporal dependence of the electric and magnetic fields is selected in the form $\exp(-i\omega t)$, so that in what follows $\mathbf{E}$ and $\mathbf{H}$ stand for the corresponding complex amplitudes. We suppose $\varepsilon''\geq 0$, i.e., we do not discuss a scatterer with a population inversion, though the obtained results can be readily generalized to this case by the formal {sign} change of $\varepsilon''$. For the sake of simplicity, we consider only the normal incidence. We choose the conventional orientation of the coordinate system, whose $z$-axis coincides with the axis of the cylinder and the wave vector $\mathbf{k}$ of the incident wave is antiparallel to {the} $x$-axis~\cite{Bohren1998}. The mutual orientations of the cylinder, the coordinate frame, and {the} vectors $\mathbf{k}$, $\mathbf{E}$, and $\mathbf{H}$ in the incident wave {are shown in} Fig.~\ref{fig:Orientation}. In this case, the incident wave fields $\mathbf{E}$ and $\mathbf{H}$ have the following components in the Cartesian coordinate system: \mbox{$\mathbf{E}=E_0(0,\sin\alpha,\cos\alpha)$;} \mbox{$\mathbf{H}=H_0(0,\cos\alpha,-\sin\alpha)$;}

We discuss the conventional real Poynting vector, whose definition (in Gaussian units) is
\begin{equation}\label{eq:Poynting_EH}
  \mathbf{S}=\frac{c}{16\pi}(\mathbf{E}^*\!\times\mathbf{H} +\mathbf{E}\times\mathbf{H}^*),
\end{equation}
where the asterisk stands for the complex conjugation.

However, sometimes the complex Poynting vector \mbox{$\hat{\mathbf{S}} = \frac{c}{16\pi} \mathbf{E}^*\!\times\mathbf{H}$} is also meaningful. Its imaginary part characterizes the alternating flow of the so-called ``stored energy''~\cite{Jackson1998}. This imaginary part plays an important role in some problems of light-matter interaction \cite{bliokh2014magnetoelectric,Bliokh2014,Bekshaev2015,Xu2019,Khonina2021,Tang2010,Lininger2022}. In most cases discussed below, the generalization of the obtained results to the complex $\hat{\mathbf{S}}$ is quite straightforward.

{In the theoretical description of the problem, the field outside the scattering object is usually presented as the sum of the incident wave field and the one scattered by the object. To avoid misunderstandings, we emphasize that anywhere in the present paper, the fields $\mathbf{E}$ and $\mathbf{H}$ in Eq.~\eqref{eq:Poynting_EH} mean the full fields equal to the indicated sum.}

To make the results invariant against the specific choice of the system of units (Gaussian, SI, etc.), we normalize $\mathbf{E}$, $\mathbf{H}$, $\mathbf{S}$, and radius vector $\mathbf{r}$ on the corresponding values in the incident wave and the radius of the cylinder's normal cross section ($R$), respectively. In what follows, we employ only these dimensionless quantities. We keep for them the same notations since it cannot give rise to confusion.

All specific examples are based on the exact solution of the problem. This solution exists at an arbitrary orientation of the vector $\mathbf{k}$ and polarization plane against the cylinder axis. However, in the general case, it is very cumbersome~\cite{Bohren1998}. Therefore, for the problem in question, it is more convenient, employing its linearity, to present the incident wave fields $\mathbf{E}_{\rm inc}$ and $\mathbf{H}_{\rm inc}$ as a sum of the fields of the pure TM- and TE-polarized waves, namely \mbox{$\mathbf{E}_{\rm inc}=\mathbf{E^{\text{\tiny
TM}}_{\rm inc}}+\mathbf{E^{\text{\tiny TE}}_{\rm inc}}$;} \mbox{$\mathbf{H}_{\rm inc}=\mathbf{H^{\text{\tiny TM}}_{\rm inc}}+\mathbf{H^{\text{\tiny TE}}_{\rm inc}}$} and use the more simple exact solutions for the pure TM and TE polarizations~\cite{Bohren1998}. Here the vectors $\mathbf{E^{\text{\tiny TM,\/TE}}_{\rm inc}}$ and $\mathbf{H^{\text{\tiny TM,\/TE}}_{\rm inc}}$ have the following components in the Cartesian coordinate system; see Fig.~\ref{fig:Orientation}:
\begin{equation}\label{eq:EHincTM}
  \mathbf{E^{\text{\tiny TM}}_{\rm inc}}=(0,0,\cos\alpha),\;\; \mathbf{H^{\text{\tiny TM}}_{\rm inc}}=(0,\cos\alpha,0),
\end{equation}
and
\begin{equation}\label{eq:EHincTE}
  \mathbf{E^{\text{\tiny TE}}_{\rm inc}}=(0,\sin\alpha,0),\;\; \mathbf{H^{\text{\tiny TE}}_{\rm inc}}=(0,0,-\sin\alpha).
\end{equation}
Then, the fields $\mathbf{E}$ and $\mathbf{H}$, {both} outside and inside the cylinder, are equal to the sum of those given by the solutions for the pure TM- and TE-polarized incident waves, whose amplitudes are defined above. Next, knowing the fields $\mathbf{E}$ and $\mathbf{H}$, we calculate $\mathbf{S}$ according to Eq.~\eqref{eq:Poynting_EH}.

\section{Symmetry\label{sec:symmetry}}

For the pure TM or TE polarizations, the problem is symmetric against the transformation \mbox{$z\rightarrow -z$.} Accordingly, the plane \mbox{$z=0$} is invariant for the Poynting vector field lines. An oblique orientation of the polarization plane with respect to the $z$-axis breaks this symmetry, even if the vector $\mathbf{k}$ still remains perpendicular to the $z$-axis. As a result, the plane \mbox{$z=0$} {\it is not invariant anymore}: generally speaking, for the points belonging to this plane, the $z$-component of the Poynting vector ($S_z$) does not vanish; see also the last paragraph of this section. However, the problem still possesses the translational symmetry along the $z$-axis. It means, that the fields $\mathbf{E}$, $\mathbf{H}$, and $\mathbf{S}$ depend on the two spatial variables $x$ and $y$ solely.

To understand more subtle properties of the symmetry, note that owing to the linearity of the problem, different components of fields  $\mathbf{E}$ and $\mathbf{H}$ are coupled only through the boundary conditions at the surface of the cylinder. These conditions stipulate the continuity of the tangential components of the fields. How does it affect the scattered fields outside the cylinder and the ones inside it for the pure TM- or TE-polarized incident wave? Consider, for definiteness, the pure TM case. Selecting the unit vector $\mathbf{e}_z$ as one of the two independent tangential {vectors} to the cylinder surface and taking the tangential unit vector belonging to the $xy$ plane as the other (actually, this is vector $\mathbf{e}_\varphi$ of the corresponding cylindrical coordinate system; see Fig.~\ref{fig:er_ephi}), we obtain that $\mathbf{E^{\text{\tiny TM}}_{\rm inc}}$ is tangential to the cylinder surface at any point and does not have any projection onto $\mathbf{e}_\varphi$. Therefore, the $x$ and $y$ components of the corresponding scattered field and the field within the cylinder {\it are not coupled\/} {to} the incident wave {field}, i.e., they {\it are not excited}. Thus, {for all these fields, only $z$-component does not vanish}.

\begin{figure}
  \centering
  \includegraphics[width=0.8\columnwidth]{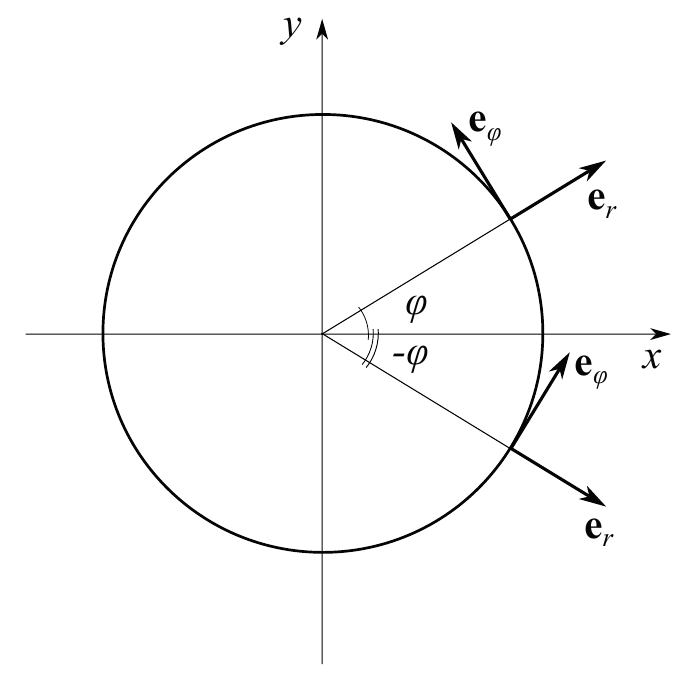}
  \caption{Unit vectors $\mathbf{e}_r$ and $\mathbf{e}_\varphi$ for an arbitrary angle $\phi$. }\label{fig:er_ephi}
\end{figure}

In contrast, $\mathbf{H^{\text{\tiny TM}}_{\rm inc}}$ does not have any projection onto  $\mathbf{e}_z$ but at any $x \neq 0$ has non-zero projections onto $\mathbf{e}_\varphi$, while $\mathbf{e}_\varphi$ itself has non-zero $x$- and $y$-components (see Fig.~\ref{fig:er_ephi}). Accordingly, only  $x$- and $y$-components of $\mathbf{H}$ are excited in this case. The conclusion is that for the pure TM-polarized incident wave {both} the scattered electromagnetic field and the one inside the cylinder {occur the TM polarized too}. Obviously, the analogous conclusion is valid for the pure TE polarization of the incident wave.

Then, the solution of Maxwell's equations for the given problem should be as follows:
\begin{equation}\label{eq:EHcomponentsMixedTM-TE}
  \mathbf{E}=(E^{\text{\tiny TE}}_x,E^{\text{\tiny TE}}_y,E^{\text{\tiny TM}}_z),\;\; \mathbf{H}=(H^{\text{\tiny TM}}_x,H^{\text{\tiny TM}}_y,H^{\text{\tiny TE}}_z),
\end{equation}
where the superscripts TM and TE designate the solutions for the corresponding pure-polarized incident waves, whose amplitudes are given by Eqs.~\eqref{eq:EHincTM} and \eqref{eq:EHincTE}, respectively. Importantly, all components of $\mathbf{E}$ and $\mathbf{H}$ in Eq.~\eqref{eq:EHcomponentsMixedTM-TE} are functions of $x$ and $y$ and do not depend on~$z$. Is there any {additional} symmetry in this coordinate dependence?

According to the made problem formulation, the incident wave propagates against the $x$-axis. It violates the $x\rightarrow -x$ symmetry. What is about the symmetry against the $y\rightarrow -y$ transformation? To answer the question, note that in the cylindrical coordinate system, the transformation $y\rightarrow -y$ is equivalent to $\varphi\rightarrow -\varphi$; see Fig.~\ref{fig:er_ephi}. This transformation does not affect the \mbox{$y$-component} of the $\mathbf{e}_\varphi$ equal to $\cos\varphi$ and changes the sign of \mbox{$(e_\varphi)_x = \sin\varphi$} to the opposite; see Fig.~\ref{fig:er_ephi}. Accordingly, for the pure TM or TE polarizations both, the $y$-components of the fields $\mathbf{E}$ and $\mathbf{H}$ (if any) should be invariant against this transformation, while their \mbox{$x$-components} should change the signs. Regarding the $z$-components,
the transformation does not act on them. Applying this rule to Eq.~\eqref{eq:EHcomponentsMixedTM-TE} we obtain
\begin{eqnarray}
  & & (E^{\text{\tiny TE}}_x(x,y),E^{\text{\tiny TE}}_y(x,y),E^{\text{\tiny TM}}_z(x,y)) = \nonumber\\
  & & (-E^{\text{\tiny TE}}_x(x,-y),E^{\text{\tiny TE}}_y(x,-y),E^{\text{\tiny TM}}_z(x,-y)), \label{eq:E_y->-y}\\
  & & (H^{\text{\tiny TM}}_x(x,y),H^{\text{\tiny TM}}_y(x,y),H^{\text{\tiny TE}}_z(x,y)) =\nonumber \\
  & & (-H^{\text{\tiny TM}}_x(x,-y),H^{\text{\tiny TM}}_y(x,-y),H^{\text{\tiny TE}}_z(x,-y)). \label{eq:H_y->-y}
\end{eqnarray}

Finally, according to Eq.~\eqref{eq:Poynting_EH}, it gives rise to the following rules for the Poynting vector components:
\begin{eqnarray}
  & & (S_x(x,y),S_y(x,y),S_z(x,y)) = \nonumber\\
  & & (S_x(x,-y),-S_y(x,-y),-S_z(x,-y)). \label{eq:S_y->-y}
\end{eqnarray}
Identities \eqref{eq:E_y->-y}--\eqref{eq:S_y->-y} also may be verified constructively based on the explicit form of the exact solution of the problem in question.

Note that the obtained results are in full agreement with the conclusions presented at the beginning of this section based on the problem symmetry. Indeed, recall that, according to Eq.~\eqref{eq:Poynting_EH}, vector $\mathbf{S}$ is perpendicular to both $\mathbf{E}$ and $\mathbf{H}$. On the other hand, as it follows from Eq.~\eqref{eq:EHcomponentsMixedTM-TE}, for
the pure TM polarization of the incident wave, vector $\mathbf{E}$ of the solution of Maxwell's equations is always parallel to the $z$-axis. The same is true for the pure TE polarization for vector $\mathbf{H}$. Then, in both these cases, the $z$-component of vector $\mathbf{S}$ must equal zero {\it identically}. In contrast, for an oblique orientation of the polarization plane, $S_z$ may vanish only at certain specific values of $x$ and $y$. These points with $S_z=0$ will be important for the subsequent analysis.

\section{Streamlines, true and false singularities\label{sec:steram}} 

The pattern of the Poynting vector streamlines determines the topological structure of this field. By definition, at any regular field point, the Poynting vector is tangential to the streamline passing through this point. Then, it is convenient to represent the streamlines in a parametric form, considering them as ``trajectories" of points moving in a three-dimensional (3D) space: \mbox{$\mathbf{r}=\mathbf{r}(t)$}. In this case, the ``velocity", $d\mathbf{r}/dt$ is tangential to the corresponding trajectory, i.e., parallel to $\mathbf{S}$ at the same point of it. It means that the vector $d\mathbf{r}/dt$ is proportional to $\mathbf{S}(\mathbf{r})$. We should stress that $t$ here is just a parameter. It is not the actual time, and $d\mathbf{r}/dt$ is not the actual velocity describing the propagation of the electromagnetic energy along the streamlines: multiplication of $t$ by any constant does not change the streamline shape. Therefore, the proper choice of this constant always makes it possible to convert to unity the proportionality coefficient between $d\mathbf{r}/dt$ and $\mathbf{S}(\mathbf{r})$. Then, eventually, the equation determining the streamlines takes the form
\begin{equation}\label{eq:dr/dt=S}
  \frac{d\mathbf{r}}{dt} = \mathbf{S}(\mathbf{r}).
\end{equation}

Now we have to make an important comment. It has been stressed above that for the problem in question, the Poynting vector field does not depend on $z$. However, {\it it is not the case for the streamlines}. {Indeed, consider an arbitrary regular point belonging to an arbitrary streamline. Generally speaking, at this point \mbox{$S_x \neq S_y \neq S_z \neq 0$.} Then, as it follows from Eq.~\eqref{eq:dr/dt=S}, for the streamline emerging from this point, the dependence on ``time'' $t$ of each of the three spatial coordinates is individual. This gives rise to a substantially three-dimensional shape of the given line near this point. Since the point is arbitrary, it is true for the streamline as a whole. The independence  $\mathbf{S}$ of $z$ means only that any translation of a given 3D streamline along the $z$-axis transforms it to the identical streamline, which is~3D~too.}

As mentioned above, the types and positions of singular points of Eq.~\eqref{eq:dr/dt=S} determine the global topological structure of the {field} $\mathbf{S}(\mathbf{r})$. Therefore, these points are of special interest to us. According to the general rules, the singularities are located at the points of intersections of the three null isoclines: $S_{x,y,z}=0$. However, for the problem under consideration, the field $\mathbf{S}$ depends only on $x$ and $y$ and does not depend on $z$. In this case we have {\it three\/} equations $S_{x,y,z}(x,y)=0$ for {\it two} variables $x$ and $y$. This problem is over-determined and does not have {\it any\/} solution. The only exception is the case when the condition {$S_{w}=0$} holds {\it identically\/} owing to the problem symmetry. Here
{$w$} denotes any of the coordinates $x,y,z$.

As pointed out above, we do have such a case for the pure TM or TE polarization when $S_z \equiv 0$ at any $x$ and $y$. What is about an oblique orientation of the polarization plane? In this case, we have {a singled out} plane \mbox{$y=0$}. Indeed, according to Eq.~\eqref{eq:S_y->-y}, $S_y(x,y) = -S_y(x,-y)$ and $S_z(x,y)=-S_z(x,-y)$. Then, \mbox{$S_y(x,0)=S_z(x,0)=0$,} and for all points belonging to the plane \mbox{$y=0$} the equations $S_{x,y,z}=0$ transforms into the single equation $S_x(x,0)=0$. Its roots, $x_{\rm sin}$ (if any), give the position of singularities. The conclusion is that, for the problem in question, the only true singularities may be observed either for the pure TM- or TE-polarized incident wave or (at an oblique polarization plane orientation) must belong to the plane $y=0$.

Now recall that the field $\mathbf{S}$ is $z$-independent. It means that, if at a singular point $\mathbf{S}(x_{\rm sin},y_{\rm sin})=0$, this condition holds at {\it any\/} value of $z$. Therefore, in all cases, instead of singular {\it points} we have continuous straight singular {\it lines\/} parallel to the $z$-axis. Moreover, the invariance of the problem against any translation along the $z$-axis makes one of the roots of the characteristic equation for these singularities identically equal to zero.

In addition to the discussed true singularities, there are the ones that we call ``false." The point is that we can consider projections of the actual 3D streamlines onto the $xy$ plane. These projections also play an essential role in understanding the $\mathbf{S}$-field topological structure. The streamline pattern in these projections is described by the two-dimensional (2D) version of Eq.~\eqref{eq:dr/dt=S}, where the vectors $\mathbf{r}$ and $\mathbf{S}$ have only $x$ and $y$ components. The corresponding 2D pattern also may have singularities whose coordinates are the solutions of the equations $S_{x,y}(x,y)=0$. However, we must remember that, as shown above, if the $y$ coordinate of these points is not equal to zero, $S_z$ does not vanish there. Thus, in 3D, these points remain regular.

To illustrate this general discussion, we present an example of the true and false singular points; see Fig.~\ref{fig:2D-3D}. The figure corresponds to the irradiation of a right circular cylinder with the permittivity \mbox{$\varepsilon = 17.775+0.024 i$.} It is the permittivity of germanium at $\lambda=1590$~nm~\cite{Polyanskiy}; $\alpha = 45.403^\circ$, the size parameter $kR = 1.62$. Note the sharp drop of $|\mathbf{S}|$ in the vicinity of the $x$-axis owing to the vanishing of $S_{y,z}$ at $y=0$.

\section{Phenomenology\label{sec:phenomenology}}

\begin{figure*}[ht]
  \centering
  \includegraphics[width=\textwidth]{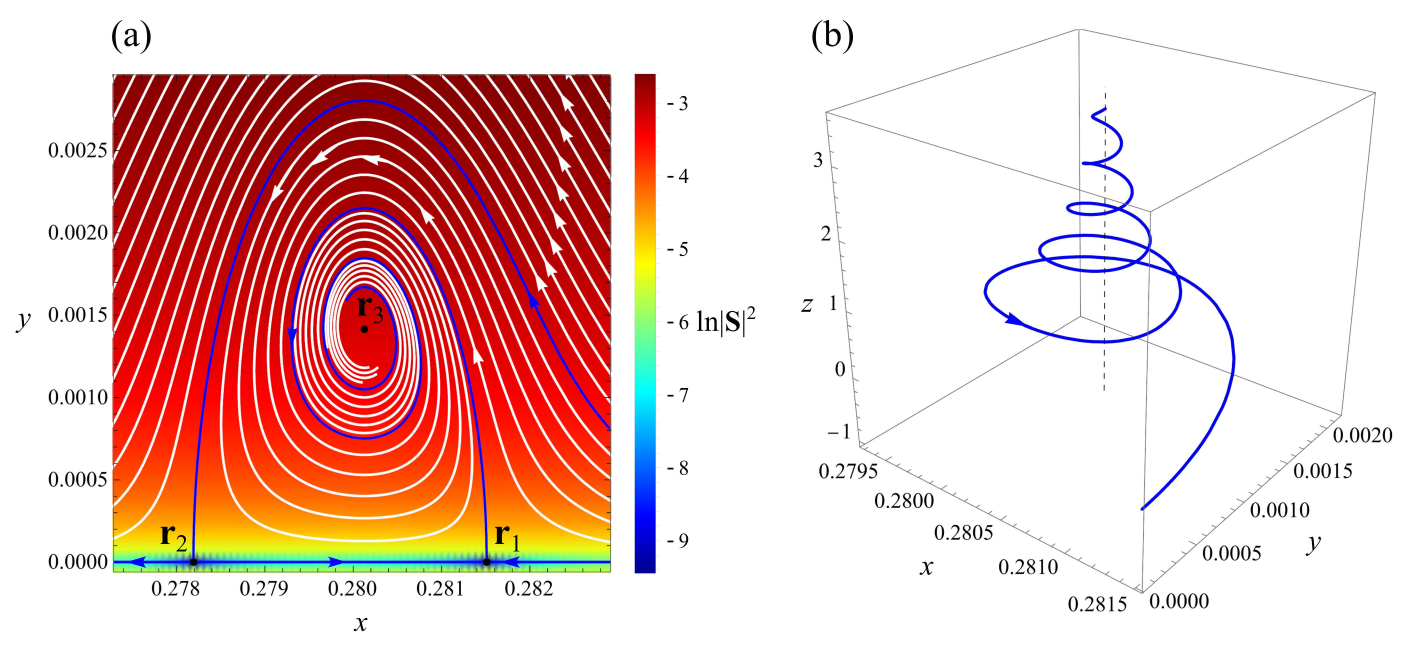}
  \caption{The Poynting vector field within a germanium circular cylinder; $\alpha = 45.403^\circ$. {The spatial coordinates are normalized on the cylinder radius $R$.} (a) 2D projection of the streamline pattern exhibits three singularities indicated as black dots: two saddles ($\mathbf{r}_{1,2}$) lying on the $x$-axis and a stable focus ($\mathbf{r}_3$) which does not belong to this axis. While the saddles are the true singularities, the focus is not. It is {seen} in (b) {showing} a 3D image of the separatrix emerging from saddle $\mathbf{r}_1$ and winding around the vertical straight line (dashed), whose projection onto the $xy$-plane corresponds to $\mathbf{r}_3$. See text for details. Pay attention to the difference in the scales of the axes. }\label{fig:2D-3D}
\end{figure*}

Here we elucidate how the problem symmetry affects bifurcations in the Poynting vector field. To make the consideration the most general, we develop a phenomenological approach, taking into account only the restrictions imposed by the symmetry, spatial dimension, and energy conservation law. We use the explicit form of the solution only to {\it illustrate \/} the general results.

The main idea is to expand the field $\mathbf{S}(x,y)$ in powers of the {deviations of} $x$ and $y$ from a certain {singled-out} point (see below). {Higher symmetry of the problem corresponds to a larger number of constraints}. Therefore, in what follows, we consider the most complicated cases of a creation (annihilation) of singularities {belonging to} the plane of the mirror symmetry, $y=0$, when the number of the constraints is maximal. To describe bifurcations occurring in less symmetric points, one has to remove the constraints which do not arise in these cases.

As shown above, for the problem in question, the only true singularities are parallel to the $z$-axis straight lines belonging to the plane $y=0$. The other ``singularities'' are false since, for them, $S_z\neq 0$. However, if we consider the projection of the streamlines onto the plane $z=0$ for the corresponding 2D pattern, the false singularities transform into true 2D singular points. We discuss the creation (annihilation) of singular points of this 2D pattern occurring in the $x$-axis or close to it.

Let $\alpha$ be a control parameter whose smooth variations determine the topological structure of the pattern. In the preceding sections, it was the angle between the vector $\mathbf{E}$ of the incident wave and the $z$-axis. However, it could be {any other} quantity, e.g., the size parameter, the real, as well as the imaginary part of permittivity, etc. Its specific definition is not essential for the {developed} approach. It is important only that there {exists} a bifurcation value of this parameter $\alpha_c$ {at which} two singularities merge and disappear or are born and {separate}.

We are interested in the values of $\alpha$ in the vicinity of $\alpha_c$. In this case, it is convenient to introduce the {\it normalized\/} control parameter
\begin{equation}\label{eq:eta}
  \eta = \pm \frac{\alpha-\alpha_c}{\alpha_c},
\end{equation}
where the sign in Eq.~\eqref{eq:eta} is selected {in such a way} that the singularities exist for positive values of $\eta$.

For the reduced 2D problem, the coordinates of singular points are the solutions of the set of equations $S_x(x,y)=0$ and $S_y(x,y)=0$. In a generic case, $S_{x,y}(x,y)$ are smooth, differentiable functions. Then, the condition $S_x(x,y)=S_x(x,-y)$, see Eq.~\eqref{eq:S_y->-y}, gives rise to the equality $\partial S_x/\partial y = 0$ at $y=0$.

Next, as shown above, $S_y(x,0)=0$. Since the singularities are the intersection points of two null isoclines, close to the bifurcation ``moment'' the line $S_x(x,y)=0$ intersects the $x$-axis (the other null isocline) in two points situated close to each other. That is to say, the functions $S_x(x,0)$ vanishes at two close points. Then, in between, there is its extremum,  where $\partial S_x/\partial x = 0$. Thus, we have obtained that, close to the bifurcation, there is a point on the $x$-axis where \mbox{$\partial S_x/\partial x = \partial S_x/\partial y = 0$}. It is precisely the point where the expansion of $\mathbf{S}(x,y)$ should be performed. In the subsequent analysis, it is convenient to introduce a local coordinate frame {with the origin coinciding} with this point. The point itself may be either an extremum of the function $S_x(x,y)$ or its saddle point. We discuss these two cases separately.

\subsection{Extrema of $S_x$\label{sec:extrima}}

The expansion of a smooth function about the point of its extremum in the origin of the local coordinate frame gives rise to the following {generic} expression for~$S_x(x,y)$:
\begin{equation}
  S_x \approx S_{x0} + \frac{1}{2}\left[\left(\frac{\partial^2S_x}{\partial x^2}\right)_{\!\!0}\!\!x^2 + \left(\frac{\partial^2S_x}{\partial y^2}\right)_{\!\!0}\!\!y^2\right] +\left(\frac{\partial^2S_x}{\partial x\partial y}\right)_{\!\!0}\!\!xy,
\end{equation}
where the subscript ``0'' means that the corresponding quantities are calculated at the origin of the local coordinate frame. The quantities $\left({\partial^2S_x}/{\partial x^2}\right)_{\!0}$ and $\left({\partial^2S_x}/{\partial y^2}\right)_{\!0}$ are negative, when $S_{x0}$ is a point of a local maximum of $S_x(x,0)$, and positive, if it is a local minimum. Note that since close to the point $x=0$, there are two points where $S_x(x,0)=0$, in the case of a local maximum, $S_{x0}>0$, while, in the case of a local minimum, it is negative.

The case under consideration is not generic since $S_x(x,y)$ must satisfy the condition  $S_x(x,y)= S_x(x,-y)$. It stipulates the vanishing of $\left({\partial^2S_x}/{\partial x\partial y}\right)_{\!0}$. Eventually,
\begin{equation}\label{eq:Sx_extr}
  S_x \approx \pm\left(a_0^2 - a_{xx}^2x^2- a_{yy}^2y^2\right),
  \end{equation}
where the signs plus and minus correspond to the vicinity of a local maximum and minimum of $S_x(x,y)$, respectively. Here
\begin{equation}\label{eq:ax2_ay2}
 a_0^2 \equiv |S_{x0}|,\;\! a_{xx}^2 \equiv\frac{1}{2}\left|\left(\frac{\partial^2S_x}{\partial x^2}\right)_{\!\!0}\right|,\;\!  a_{yy}^2 \equiv\frac{1}{2}\left|\left(\frac{\partial^2S_x}{\partial y^2}\right)_{\!\!0}\right|,
\end{equation}
and all $a$'s are supposed to be non-negative. In this case, the equation of the null isocline $S_x(x,y)=0$ describes an ellipse with diameters \mbox{$d_x = 2a_0/a_{xx}$} and  \mbox{$d_y = 2a_0/a_{yy}$}.

What is about the other null isocline, $S_y(x,y)=0$? Once again, we have to begin with the generic expansion of a smooth function of $x$ and $y$, taking into account all terms up to the second order in small $x$ and $y$ and imposing the constraints stipulated by the problem symmetry. Note that, according to this constraint, $S_y(x,0)=0$ at any $x$ (see above), i.e., one of the branches of the null isocline  $S_y(x,y)=0$ is the straight line $y=0$ (the $x$-axis). To satisfy this condition, all terms in the expansion of  $S_y(x,y)$ must have $y$ as a common factor. The most general type of this expression is
\begin{equation}\label{eq:Sy_general}
  S_y \approx y\left[\left(\frac{\partial S_y}{\partial y}\right)_{\!\!0} + \left(\frac{\partial^2S_x}{\partial x\partial y}\right)_{\!\!0}\!\!x +\frac{1}{2} \left(\frac{\partial^2S_x}{\partial y^2}\right)_{\!\!0}\!\!y\right]
\end{equation}

However, we have another constraint: $S_y(x,y)=-S_y(x,-y)$. It requires the vanishing of $\left({\partial^2S_x}/{\partial y^2}\right)_{\!0}$ in the above expression. Finally, we obtain
\begin{equation}\label{eq:Sy_final}
  S_y \approx y(b_{y} + x b_{xy}),
\end{equation}
where the meaning of {$b_{y,xy}$} follows from the comparison of Eq.~\eqref{eq:Sy_final} with Eq.~\eqref{eq:Sy_general}. Then, according to Eq.~\eqref{eq:Sy_final} the null isocline $S_y=0$ consists of the two perpendicular straight lines: $y=0$ and $x=-b_{y}/b_{xy}$. In this case the null isoclines $S_{x,y}=0$ have four intersection points with the coordinates
\begin{eqnarray}
  & & \mathbf{r}_{1,2} = \left(\pm \frac{a_0}{a_{xx}},0\right), \label{eq:r12} \\
  & & \mathbf{r}_{3,4}=\left(-\frac{b_{y}}{b_{xy}}, \pm \sqrt{\frac{a_0^2b_{xy}^2-a_{xx}^2b_{y}^2}{a_{yy}^2b_{xy}^2}}\right), \label{eq:r34}
\end{eqnarray}
see Fig.~\ref{fig:ellipse}
\begin{figure}
  \centering
  \vspace*{6pt}
  \includegraphics[width=.8\columnwidth]{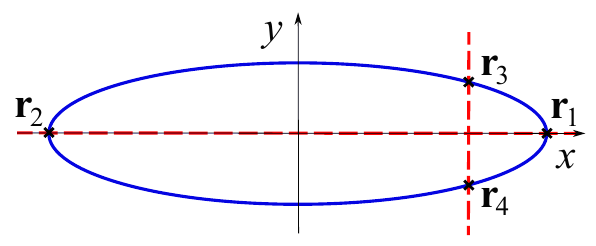}
  \caption{Null isocline $S_x=0$ (blue ellipse, solid line) and two branches of null isocline $S_y=0$ (red dashed lines). Singular points $\mathbf{r}_{1,2,3,4}$ are marked with crosses. When the control parameter $\eta$ decreases, the ellipse contracts, while the dashed vertical line remains at the same position. As a result, point $\mathbf{r}_{1}$ moves along the $x$-axis toward the vertical branch of null isocline $S_y=0$, while points  $\mathbf{r}_{3,4}$ move toward the $x$-axis. Eventually, these three points merge. For the smaller values of $\eta$ only $\mathbf{r}_{1,2}$ remain. With a further decrease in $\eta$ and the corresponding contraction of the ellipse, $\mathbf{r}_{1,2}$ {approach} each other and merge at $\eta=0$. See text for details.}\label{fig:ellipse}
\end{figure}

{\bf Importantly}, all coefficients $a$ and $b$ in the above expressions depend on the normalized control parameter, $\eta$. Moreover, at $\eta = 0$, two points $\mathbf{r}_{1,2}$ merge at the origin of the coordinate frame (we remind that the origin is at the point of the local extremum, which lies in between $\mathbf{r}_{1,2}$). At this ``moment,'' the origin inherits the singular nature of $\mathbf{r}_{1,2}$ and becomes a singular point of the $\mathbf{S}$-field. It means that $\mathbf{S}(0,0)=0$ at $\eta = 0$. Then, expanding $S_{x0}$ in powers of small $\eta$, we obtain that the expansion of $a_0^2\equiv |S_{x0}|$ (not the one for $a_0(\eta)$!) begins from a linear in $\eta$ term.

Regarding all other $a$'s and $b$'s, in a generic case, their expansions begin with constants. Then, at small $\eta$ tending to zero, the diameters of the ellipse in Fig.~\ref{fig:ellipse} contracts as $\sqrt{\eta}$, see Eq.~\eqref{eq:ax2_ay2}. At the same time, the vertical line of the null isocline $S_y=0$ remains practically fixed. If so, at $\eta \rightarrow 0$ the points $\mathbf{r}_{3,4}$ move toward $\mathbf{r}_1$ (or $\mathbf{r}_2$, depending on the sign of $-b_{y}/b_{xy}$); eventually, merge with it and annihilate.

Thus, a decrease in $\eta$ results in two sequential bifurcations corresponding to the fusion of the singularities. The first bifurcation is $3 \rightarrow 1$ ({coalescence} of $\mathbf{r}_{3,4}$ with {either} $\mathbf{r}_1$ or $\mathbf{r}_2$), the second is $2\rightarrow 0$ ({fusion} of $\mathbf{r}_1$ {and} $\mathbf{r}_2$).

However, this scenario has an exception, namely the case $b_{y}=0$. In this case, the points $\mathbf{r}_{3,4}$ are situated on the $y$-axis. Then, instead of the two sequential bifurcations, we have a single bifurcation of the type $4\rightarrow 0$.

To unveil the physical grounds for vanishing $b_{y}$, we have to recall the energy conservation law. For the problem under consideration, its differential form reads~\cite{Landau2013}:
\begin{equation}\label{eq:divS=-q}
  {\rm div}\, \mathbf{S} = -q,
\end{equation}
where $q$ is the {power} dissipated in unit of volume. In our case, ${\rm div}\,\mathbf{S} = \partial S_x/\partial x + \partial S_y/\partial y$ since $\partial S_z/\partial z =0$. Then, according to Eqs.~\eqref{eq:Sx_extr} and \eqref{eq:Sy_final} we obtain that in the given approximation
\begin{equation}\label{eq:divS_extr}
  b_{y}+\left(b_{xy} \mp 2a_{xx}^2\right)x=-q.
\end{equation}

In non-dissipative media $q\equiv 0$, i.e., the l.h.s. of Eq.~\eqref{eq:divS_extr} must vanish {\it identically.} It results in the conditions
\begin{equation}\label{eq:by1=0_nondiss}
  b_{y}=0,\;\; b_{xy} = \pm 2a_{xx}^2.
\end{equation}
Thus, based just on the general requirements following from the problem symmetry and energy conservation, we have obtained that the sequence of the bifurcations $3 \rightarrow 1$ and  $2 \rightarrow 0$ may take place in dissipative media only, while non-dissipative ones may exhibit the degenerate bifurcation  $4 \rightarrow 0$ solely.

To understand the types of the {fusing} singular points
{we have to consider the values of the Jacobian matrix of the system of equations $S_{x,y}(x,y)=0$ at every singular point $\mathbf{r}_n$}
and find {its eigenvalues, i.e.,} the roots $\kappa_n$ of the corresponding characteristic equation.

The degenerate non-dissipative case, when $q=0$ and constraint~\eqref{eq:by1=0_nondiss} is applicable, is more straightforward for the analysis, so we begin with it. In this case, applying the above procedure to \mbox{Eqs.~\eqref{eq:Sx_extr}--\eqref{eq:r34}} gives rise to the following expressions valid in the proximity of both local maxima and minima of $S_x(x,y)$:
\begin{equation}\label{eq:kappa12_nondiss}
\kappa^{\pm}_1 = \kappa^{\pm}_{2}= \pm 2 a_0 a_{xx}, \;\; \kappa^{\pm}_{3}=\kappa^{\pm}_{4}=\pm 2i a_0 a_{xx},
\end{equation}
{where the subscripts have the same meaning as those in $\mathbf{r}_{1,2,3,4}$} (see Fig.~\ref{fig:ellipse}).
Thus, points $\mathbf{r}_{1,2}$ are saddles, while $\mathbf{r}_{3,4}$ are centers.

Note, $\mathbf{r}_i$ and $\kappa_{i}^{\pm}$ ($i=1,2,3,4$)
all are scaled as $\sqrt{\eta}$ owing to Eqs.~\eqref{eq:r12},~\eqref{eq:r34},~\eqref{eq:by1=0_nondiss}, \eqref{eq:kappa12_nondiss}, and the mentioned above $\eta$-dependence of coefficients $a$'s, $b$'s. In other words, here we have a {typical} pitchfork bifurcation.

Next, for the discussed 2D streamline pattern the conditions stipulated by Eq.~\eqref{eq:S_y->-y} mean that the pattern
possesses a mirror symmetry with respect to the $x$-axis. Then, the reflection against this axis transforms the stable and unstable separatrixes of the saddles $\mathbf{r}_{1,2}$ into themselves. However, there are just two types of lines remaining invariant {for} this transformation, namely the $x$-axis itself and any line parallel to the $y$-axis. Thus, for the saddles {belonging to} the $x$-axis, one of the separatrixes is the $x$-axis, while the other is {normal} to it. Moreover, since a certain bunch of streamlines, {emerging} from the vicinity {of} one saddle along the unstable direction, {approaches the} proximity of the other along its stable direction, it means that if for one saddle the separatrix parallel to the $y$-axis is unstable (the corresponding $\kappa$ is positive), for the other it is stable ($\kappa <0$). The same conclusion is valid for the separatrixes coinciding with the $x$-axis. This result may also be verified by directly calculating the eigenvectors {at} these singular points.

What happens in dissipative media? To {answer} the question, we have to discuss the explicit expression for $q$ in Eq.~\eqref{eq:divS=-q}. Bearing in mind that, for the problem under consideration, permeability $\mu \equiv 1$, in the {chosen} dimensionless variables
{one has}
\begin{equation}\label{eq:q}
  q = \varepsilon''kR|\mathbf{E}|^2,
\end{equation}
where $k=\omega/c$ is the wavenumber of the incident wave and $c$ stands for the speed of light in a vacuum~\cite{Landau2013,Tribelsky2022_Nanomat_diss}.

Now recall that Eqs.~\eqref{eq:Sx_extr},~\eqref{eq:Sy_final} for $S_x$ and $S_y$ are approximate: only {the terms} quadratic in both $x$ and $y$ are taken into account. Accordingly, the accuracy of {\rm div}$\,\mathbf{S}$, calculated based on these expressions, is up to linear terms in $x$ and $y$. To maintain the same accuracy in the expansion for $q(x,y)$, we must drop all terms higher than linear {ones}.

On the other hand, as it follows from Eq.~\eqref{eq:E_y->-y}, $E_x(x,0)=0$. Then, the expansion of $E_x(x,y)$ in powers of small $x$ and $y$ begins with a linear term, and hence $|E_x|^2$ is quadratic in small $y$ and should be dropped.

Next, the same Eq.~\eqref{eq:E_y->-y} requires that $\partial E_y/\partial y =  \partial E_z/\partial y =0$ at $y=0$. It means that linear in $y$ terms do not enter the expansions of $E_{y,z}(x,y)$. Regarding the quadratic terms, they should be neglected according to the specified accuracy. Thus, eventually, we obtain
\begin{equation}\label{eq:q_approx}
  q \approx  \varepsilon''kR\left(|E_y(0,0)|^2 + |E_z(0,0)|^2\right) = const >0.
\end{equation}
In this case, to keep the equality in Eq.~\eqref{eq:divS_extr} valid at any $x$ we must change Eq.~\eqref{eq:by1=0_nondiss} to the following
\begin{equation}\label{eq:by1=const_diss}
  b_{y} \approx const <0,\;\; b_{xy} \approx \pm 2a_{xx}^2.
\end{equation}
Note that the condition $b_{xy} \mp 2a_{xx}^2 \approx 0$ removes the problem {of} the non-physical negativeness of $q$ at \mbox{$x < b_y/(2a_{xx}^2\mp b_{xy})$}, which otherwise would arise according to Eq.~\eqref{eq:divS_extr}.

In what follows, for definiteness, we consider the vicinity of a local maximum of $S_x(x,y)$, {choosing} {the plus sign on} the r.h.s. of Eq.~\eqref{eq:Sx_extr} (the vicinity of a local minimum is {considered} in the same manner). In this case, $ b_{xy} \approx 2a_{xx}^2 >0$ and {the} points $\mathbf{r}_{3,4}$ {belong to the right semiplane}, as shown in Fig.~\ref{fig:ellipse}.

Then, the same approach as that described above for the non-dissipative case gives rise to the following results
\begin{eqnarray}
& & \kappa^{+}_{1,2} = \pm 2a_0a_{xx}+b_y,\;\; \kappa^{-}_{1,2} = \mp 2a_0a_{xx}, \label{eq:kappa12_diss}\\
& & \kappa^{\pm}_{3} =\kappa^{\pm}_{4} = \left(b_y \pm \sqrt{5b_y^2-(4a_0 a_{xx})^2}\;\right)/2. \label{eq:kappa34_diss}
\end{eqnarray}
In Eq.~\eqref{eq:kappa12_diss},
the upper and lower signs correspond to $\mathbf{r}_1$ and $\mathbf{r}_2$, respectively. Note, as follows from Eqs.~\mbox{\eqref{eq:kappa12_diss}--\eqref{eq:kappa34_diss},} $\kappa_i^{+}+\kappa_i^{-} = b_y$; $i=1,2,3,4.$

{Consider} the obtained eigenvalues. We remind that, for the case under consideration, $b_y<0$, while all other $a$'s and $b$'s are nonnegative; and, for the approximation made, the only coefficient depending on the control parameter $\eta$ is $a_0 \propto \sqrt{\eta}$. All other coefficients are supposed to be constants. We will move from a certain initial {positive} value of $\eta$ toward $\eta = 0$ so that $a_0(\eta)$ decreases monotonically.

As for point $\mathbf{r}_2$, the change of $\eta$ does not affect its type. Indeed, according to Eq.~\eqref{eq:kappa12_diss},
{\mbox{$\kappa^{+}_2 = -2a_0a_{xx}+b_y <0$;}  \mbox{$\kappa^{-}_2 = 2a_0a_{xx}>0$.} }
These $\kappa$'s keep the same sign at any value of $a_0(\eta)>0$, so the point  $\mathbf{r}_2$ always is a saddle.

Regarding point $\mathbf{r}_1$, the eigenvalue {$\kappa^{-}_{1} =-2a_0a_{xx}$}
remains a real negative quantity at any $a_0$. In contrast,
{$\kappa^{+}_{1}=2a_0a_{xx}+b_y $}
changes its sign as $a_0$ decreases. As for eigenvalues $\kappa_{3,4}$, characterizing singularities $\mathbf{r}_{3,4}$, they also change their type, when $a_0$ varies. Specifically, there are three critical values of $a_0(\eta)$, namely \mbox{(i) $a_0 = -b_y\sqrt{5}/(4a_{xx})$,} \mbox{(ii) $a_0 = -b_y/(2a_{xx})$} and \mbox{(iii) $a_0=0$}. Let us discuss the streamline pattern in the proximity of each of them.

\subsubsection{$S_{x0}\equiv a_0^2 > 5b_y^2/(4a_{xx})^2$\label{sec:5.1.1}} 

Here the expression under the square root in Eq.~\eqref{eq:kappa34_diss} is negative. Then, $\kappa_{3,4}$ are complex quantities with the same negative real part, i.e., the points $\mathbf{r}_{3,4}$ are stable foci. Regarding
{$\kappa^{+}_{1}$}, for these values of $a_0$ it is positive, so here the point $\mathbf{r}_1$ is a saddle.

\subsubsection{$S_{x0}\equiv a_0^2 = 5b_y^2/(4a_{xx})^2$\label{sec:5.1.2}} 

Here the square root in Eq.~\eqref{eq:kappa34_diss} vanishes,
{$\kappa^{\pm}_{3,4}$}
become real negative quantities, and foci $\mathbf{r}_{3,4}$ transform into stable nodes. As for singularity $\mathbf{r}_1$, at this value of $a_0$ we have
\mbox{$\kappa^{+}_1 = 2a_0a_{xx}(1-2/\sqrt{5})>0$}. Thus, $\mathbf{r}_1$ still is a saddle.

\subsubsection{$ 5b_y^2/(4a_{xx})^2 >S_{x0}\equiv a_0^2> b_y^2/(2a_{xx})^2$\label{sec:5.1.3}} 

Here we have
{$\kappa^{\pm}_{3,4}<0$} so points $\mathbf{r}_{3,4}$ remain stable nodes. Regarding, $\mathbf{r}_1$, the quantity
{$\kappa^{+}_1$} is positive in this domain, and the singularity type for $\mathbf{r}_1$ does not change: it is a saddle.

\subsubsection{$S_{x0}\equiv a_0^2=b_y^2/(2a_{xx})^2$\label{sec:5.1.4}} 

At this value of $a_0$ three points $\mathbf{r}_{1,3,4}$ merge; see Eqs.~\eqref{eq:r34},~\eqref{eq:by1=const_diss}, and Fig.~\ref{fig:ellipse}:
{$\kappa^{+}_{1,3,4}=0$, \mbox{$\kappa^{-}_{1,3,4}=-2a_0a_{xx}<0$.}}

\subsubsection{$b_y^2/(2a_{xx})^2>S_{x0}\equiv a_0^2>0$\label{sec:5.1.5}} 

In this case points $\mathbf{r}_{3,4}$
{disappear}; see Fig.~\ref{fig:ellipse}. Point $\mathbf{r}_2$ remains a saddle, while point $\mathbf{r}_1$ inherits the properties of absorbed singularities $\mathbf{r}_{3,4}$ and becomes a stable node.

\subsubsection{$S_{x0}\equiv a_0^2=0$\label{sec:5.1.6}}

Here points $\mathbf{r}_{1,2}$ merge. {At $\eta <0$}, there are no singularities in the vicinity of the origin of the local coordinate system.

It is convenient to introduce $\eta_{1,2}$ so that

\begin{equation}\label{eq:eta_n}
  a_0(\eta_1)=-\frac{b_y\sqrt{5}}{4a_{xx}},\;\; a_0(\eta_2)=-\frac{b_y}{2a_{xx}},
\end{equation}
(we remind, $b_y<0$) and the corresponding local control parameters \mbox{$\tilde{\eta}_n=(\eta -\eta_n)/\eta_n$,} ($n=1,2$). Then, expanding $a_0^2(\eta)$ in powers of $\eta$, we obtain the following full bifurcation picture.

\subsubsection{Proximity of $\eta=\eta_1$\label{sec:5.1.7}}

The number of the singularities does not change. At $\tilde{\eta}_1>0$, points $\mathbf{r}_{3,4}$ are stable foci. At $\tilde{\eta}_1 \rightarrow 0$ from the side of positive $\tilde{\eta}_1$, the imaginary parts of
{$\kappa^{\pm}_{3,4}$ }
tend to zero as $\sqrt{\tilde{\eta}_1}$. At $\tilde{\eta}_1<0$, points $\mathbf{r}_{3,4}$ are stable nodes.

\subsubsection{Proximity of $\eta=\eta_2$\label{sec:5.1.8}} 

The stable nodes $\mathbf{r}_{3,4}$, existing at $\tilde{\eta}_2>0$, merge with point $\mathbf{r}_1$ and disappear. At vanishing $\tilde{\eta}_2$,
{$\kappa^{+}_{1,3,4}$ } also vanish as
{\it linear\/} functions of $\tilde{\eta}_2$. These linear functions remain the same for both  $\tilde{\eta}_2>0$ and  $\tilde{\eta}_2<0$. In contrast, the distances between $\mathbf{r}_{1,3,4}$ are meaningful only at $\tilde{\eta}_2>0$ and vanish as $\sqrt{\tilde{\eta}_2}$ at $\tilde{\eta}_2 \rightarrow 0$; see Eqs.~\eqref{eq:r12},~\eqref{eq:r34}.

\subsubsection{Proximity of $\eta=0$\label{sec:5.1.9}} 

Stable node $\mathbf{r}_1$ merges with saddle $\mathbf{r}_2$ at $\eta = 0$ and
{both} annihilate. Before the annihilation {$\kappa^{-}_{1,2}$} have the same modulus but opposite signs and vary as $\sqrt{\eta}$; see Eq.~\eqref{eq:kappa12_diss}. The distance between the two singularities also obeys the same square root law; see Eq.~\eqref{eq:r12}. Note also that {$\kappa^{-}_2>0$}, see Eq.~\eqref{eq:kappa12_diss} and the corresponding unstable separatrix of saddle $\mathbf{r}_2$ is directed to the stable node  $\mathbf{r}_1$ along the $x$-axis.

\subsection{Saddles of $S_x$\label{sec:saddles}}

In the vicinity of a saddle point of the surface $S_x(x,y)$, the coefficients {of} $x^2$ and $y^2$ in Eq.~\eqref{eq:Sx_extr} have opposite signs, while Eq.~\eqref{eq:Sy_final} remains the same. However, this seemingly minor change in the governing equations gives rise to drastic changes in the bifurcation scenario. Let us discuss it in detail. For definiteness, we suppose that instead of Eq.~\eqref{eq:Sx_extr} now we have
\begin{equation}\label{eq:Sx_saddle}
S_x \approx S_{x0} - a_{xx}^2x^2 + a_{yy}^2y^2,
\end{equation}
where, as before, $a_{xx,yy}>0$, however, now the sign of $S_{x0}$ {is not fixed}. The case
\begin{equation}
 S_x \approx S_{x0} + a_{xx}^2x^2 - a_{yy}^2y^2,
\end{equation}
is quite analogous and will not be discussed.

The transition of the expression for $S_x(x,y)$ from Eq.~\eqref{eq:Sx_extr} to Eq.~\eqref{eq:Sx_saddle} does not affect  Eq.~\eqref{eq:by1=const_diss}: it keeps the same form with $b_y \approx const<0$ and $b_{xy} \approx 2a_{xx}>0$. The null isocline $S_y=0$ also does not change. However, for the null isocline $S_x=0$, instead of an ellipse, now we have hyperbolas; see Fig.~\ref{fig:hyperbolas}. It is convenient to discuss cases $S_{x0}>0$ and $S_{x0}<0$ separately.

\underline{$S_{x0}\equiv a_0^2>0$.} In this case, depending on the $a_0$  value, there are either only two singular points $\mathbf{r}_{1,2}$ on the \mbox{$x$-axis}, or two more singularities $\mathbf{r}_{3,4}$ lying symmetrically above and below the \mbox{$x$-axis}, respectively; see Fig.~\ref{fig:hyperbolas}a. Their coordinates are given by the same Eqs.~\eqref{eq:r12},~\eqref{eq:r34} after the formal replacement $a_{yy}^2\rightarrow-a_{yy}^2$, cf. Eqs~\eqref{eq:Sx_extr},~\eqref{eq:Sx_saddle}. The points $\mathbf{r}_{3,4}$ exist at $a_{xx}^2b_y^2\geq a_0^2b_{xy}^2$.

\underline{$S_{x0}\equiv -a_0^2<0$.} In this case, only singularities $\mathbf{r}_{3,4}$ remain; see Fig.~\ref{fig:hyperbolas}b. Their coordinates are given by Eq.~\eqref{eq:r34} after the replacement  $a_{yy}^2\rightarrow-a_{yy}^2$,  $a_{0}^2\rightarrow-a_{0}^2$.

Once again, we begin the analysis with the degenerate non-dissipative case when $b_y=0$ and the vertical branch of the null isocline $S_y=0$ coincides with the $y$-axis. In this case, at positive $S_{x0}$, points $\mathbf{r}_{3,4}$ do not exist. Regarding points $\mathbf{r}_{1,2}$, the corresponding $\kappa^{\pm}_{1,2}$ are given by the same Eq.~\eqref{eq:kappa12_nondiss}. Thus, these singular points are saddles. As $S_{x0}$ decreases, the points $\mathbf{r}_{1,2}$ move toward the origin of the coordinate frame. It is accompanied by a decrease in {$|\kappa^{\pm}_{1,2}|$}. At $S_{x0}=0$, $\mathbf{r}_{1,2}$ merge at the origin of the coordinate system, and {$\kappa^{\pm}_{1,2}$} vanish.

A further decrease in $S_{x0}$ makes it negative,
{and two new singular points $\mathbf{r}_{3,4}$ emerge from the origin and separate, symmetrically moving along the $y$-axis in opposite directions.}

\begin{figure}
  \centering
  \includegraphics[width=\columnwidth]{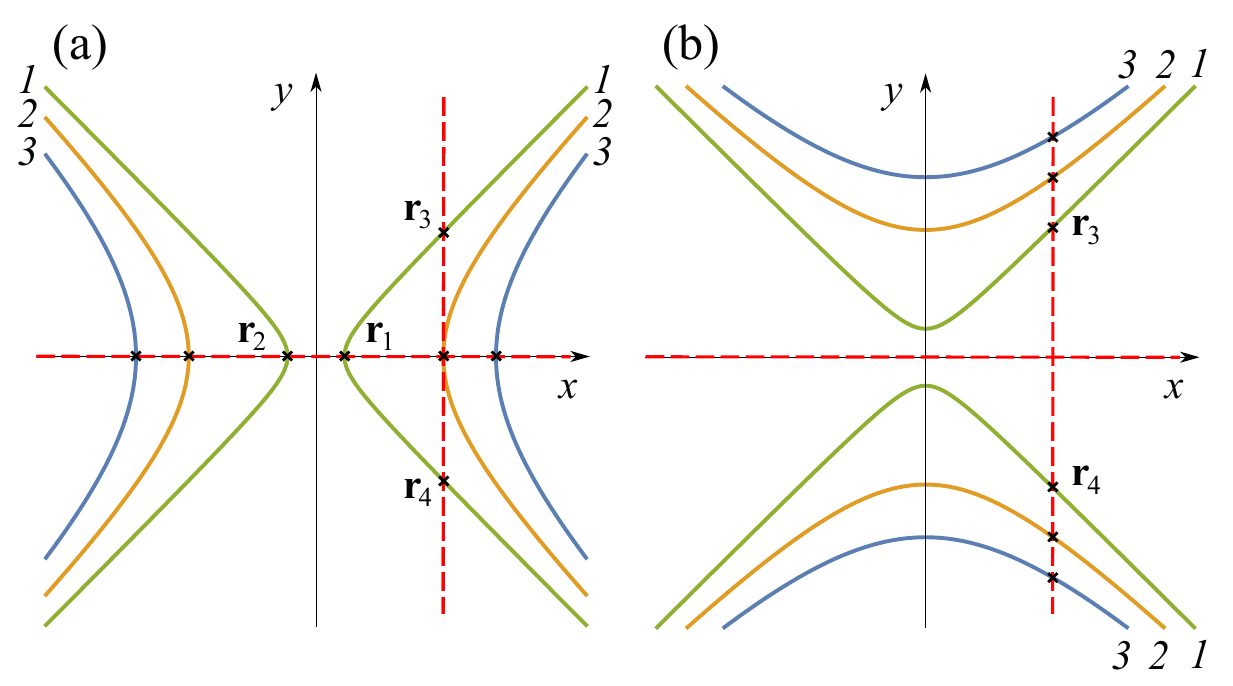}
  \caption{Null isoclines $S_x(x,y)=0$ (solid) and \mbox{$S_y(x,y)=0$} (dashed) in the vicinity of a saddle point of the surface $S_x(x,y)$, at different values of $S_{x0}$. (a) $S_{x0}>0$; (b) $S_{x0}<0$. In the {discussed} approximation, {$S_{x0}$} changes do not affect the branches of the null isocline \mbox{$S_y(x,y)=0$} but do the ones of \mbox{$S_x(x,y)=0$}. Numbers 1, 2, and 3 mark three typical positions of the latter at three different values of $S_{x0}$. An increase in the numbers corresponds to {growth in} $|S_{x0}|$. See text for details.}\label{fig:hyperbolas}
\end{figure}

Notably, in contrast to the cases discussed in the previous subsection, now these points are saddles, whose  {$\kappa^{\pm}_{3,4}$} still equal to $\pm 2a_0a_{xx}$ (we remind that for $S_{x0}<0$ the quantity $a_0$ is defined as $\sqrt{|S_{x0}|}$).

Collecting all together, we can say that, in this case, we always have two saddles situated either on the $x$- or \mbox{$y$-axis.} Their
{$\kappa^{\pm} = \pm 2a_0a_{xx}$},
and are scaled as $\sqrt{|\eta|}$. The distance between the singularities also obeys the same square root law.

This analysis shows that all equations derived in the previous subsections may be applied to the present case after the formal replacement \mbox{$a_{yy}^2\rightarrow-a_{yy}^2$,} at \mbox{$S_{x0}\geq 0$,} supplemented by the transformation \mbox{$a_{0}^2\rightarrow-a_{0}^2$,} at \mbox{$S_{x0} < 0$.} For this reason, we can proceed to the inspection of the general bifurcation picture for the dissipative case ($b_y \neq 0$) based on the results obtained in subsection~\ref{sec:extrima} in the proximity of an extremum of $S_x(x,y)$. It is convenient to begin the discussion from positive values of $S_{x0}$, gradually decreasing~it.

\subsubsection{$S_{x0}\equiv a_0^2>b_y^2/(2a_{xx})^2$\label{sec:5.2.1}} 

In this case (curves 3 in Fig.~\ref{fig:hyperbolas}a) only two singularities, $\mathbf{r}_{1,2}$ exist. They lie on the $x$-axis symmetric against the point $x=0$, and their coordinates are given by Eq.~\eqref{eq:r12}. As the control parameter varies, the distance between these points changes as $a_0$, i.e., as $\sqrt{\eta}$, where $a_0=0$ at $\eta =0$. The eigenvalues of these singularities are given by Eq.~\eqref{eq:kappa12_diss}, which is not affected by the replacement rules specified above since it does not contain $a_{yy}$ and the sign of $a_0$ remains the same as that in Eq.~\eqref{eq:kappa12_diss}, because  $S_{x0}>0$. Then, we easily obtain that  $\kappa^\pm_{1,2}$ are purely real in the discussed range of $S_{x0}$ values, and $\kappa^\pm_i$ have opposite signs at a given $i\;(i =1,2)$. That is to say, both singularities are saddles.

\subsubsection{$S_{x0}\equiv a_0^2=b_y^2/(2a_{xx})^2>0$\label{sec:5.2.2}} 

At this value of $S_{x0}$ two more singular points, $\mathbf{r}_{3,4}$ begin to detach from $\mathbf{r}_1$; see right curve 2 in Fig.~\ref{fig:hyperbolas}a. The eigenvalues of the resulting degenerate singularity are
$\kappa^+=0$ and \mbox{$\kappa^-=b_y<0$}.

\subsubsection{$b_y^2/(2a_{xx})^2 > S_{x0}\equiv a_0^2>0$\label{sec:5.2.3}} 

According to Eq.~\eqref{eq:kappa12_diss},
we have \mbox{$\kappa^+_{1,2}<0$}, \mbox{$\kappa^-_1<0$}, \mbox{$\kappa^-_2>0$}. Thus,
$\mathbf{r}_1$  is a stable node, while $\mathbf{r}_2$ is a saddle.
As for the new-born singularities, {$\mathbf{r}_{3,4}$}, their eigenvalues are given by Eq.~\eqref{eq:kappa34_diss}, which also can be applied without any modifications. For the case in question, \mbox{$5b_y^2-16a_0^2a_{xx}^2 > b_y^2>0$}. Hence, {$\kappa^{\pm}_{3,4}$} are real quantities with the opposite signs, i.e., the points $\mathbf{r}_{3,4}$ are saddles. Their coordinates are given by Eq.~\eqref{eq:r34}, where $a_0^2$ should be replace by $-a_0^2$, supplemented by the condition $b_{xy}^2=4a_{xx}^2$; see Eq.~\eqref{eq:by1=const_diss}.

As $S_{x0}$ decreases, the points $\mathbf{r}_{3,4}$ symmetrically diverge from the $x$-axis along the vertical branch of null isocline $S_y=0$, and the distance between them increases as
{$\sqrt{b_y^2-(2a_0a_{xx})^2}/(a_{yy}a_{xx})$}; see Eqs.~\eqref{eq:r34} and \eqref{eq:by1=const_diss}.
Introducing $\eta_2$, according to Eq.~\eqref{eq:eta_n}, and \mbox{$\tilde{\eta}\equiv (\eta-\eta_2)/\eta_2$} we obtain that in proximity of $\eta=\eta_2$ this distance varies as $\sqrt{\tilde{\eta}}$. In contrast,
{$\kappa^{\pm}_{3,4}$} remain linear functions of $\tilde{\eta}$.

\subsubsection{$S_{x0}=0$\label{sec:5.2.4}}

At this value of $S_{x0}$, the two branches of null isocline $S_{x}(x,y)=0$ degenerate into the two straight lines: $y=\pm (a_{xx}/a_{yy})x$ intersecting with the horizontal branch of null isocline $S_{y}(x,y)=0$ at the origin of the coordinate frame, and points $\mathbf{r}_{1,2}$ merge. According to Eq.~\eqref{eq:kappa12_diss}, the corresponding $\kappa$'s are as follows:
{$\kappa^+_{1,2}=b_y<0$ and $\kappa^-_{1,2}=0$}.

\subsubsection{$S_{x0}\equiv a_0^2<0$\label{sec:5.2.5}} 

Only singular points $\mathbf{r}_{3,4}$ remain; see Fig.~\ref{fig:hyperbolas}b. The corresponding eigenvalues equal each other and are
{$\kappa^{\pm}_{3,4} = \left(b_y \pm \sqrt{5b_y^2+16a_0^2a_{xx}^2}\,\right)/{2}$, }
cf. Eq.~\eqref{eq:kappa34_diss}. Both terms under the square root sign are positive, so these singularities are saddles.

To conclude this section, note that in the vicinity of an extremum of $S_x(x,y)$, points $\mathbf{r}_{3,4}$ can be either foci or nodes; see Sections \ref{sec:5.1.1} and \ref{sec:5.1.2}--\ref{sec:5.1.4}. This is not the case in the vicinity of its saddle. It is explained by the fact that  the eigenvalues $\kappa_{3,4}$ could be complex at \mbox{$a_0^2>5b_y^2/(4a_{xx})^2$;} see Eq.~\eqref{eq:kappa34_diss}. On the other hand, these singularities exist, provided \mbox{$a_0^2<b_y^2/(2a_{xx})^2$}; see Eq.~\eqref{eq:r34}, supplemented with the replacements \mbox{$a_{yy}^2\rightarrow -a_{yy}^2$} and $b_{xy}^2 \rightarrow 4a_{xx}^4$; see Eq.~\eqref{eq:by1=const_diss}. Then, since $1/4 < 5/16$, $\kappa_{3,4}$ remain purely real in the entire range of the existence of these singularities. That is to say, in the vicinity of a saddle point of $S_x(x,y)$, singular points $\mathbf{r}_{3,4}$ exist only as saddles.

{Finally, note that the above analysis shows that the topological structure of the Poynting vector pattern changes only as a result of a finite number of bifurcations. It means that between the points of the bifurcations, the pattern is topologically stable against small variations of the bifurcation parameter(s). The conclusion should be valid for generalizing this discussion to any continuous symmetry violation in the light scattering problem.}

\section{Examples\label{sec:examples}}

\begin{figure}
  \centering
  \includegraphics[width=\columnwidth]{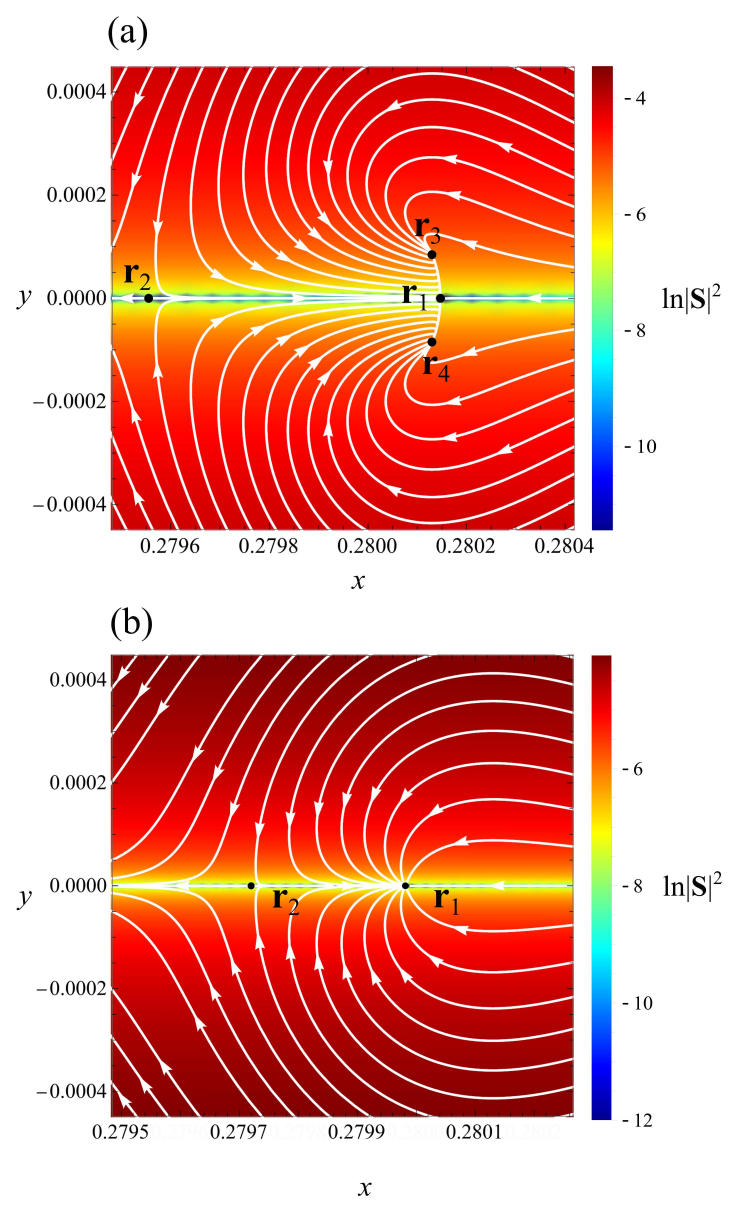}
  \caption{Projection of streamlines onto the $xy$-plane. {The spatial coordinates are normalized on the cylinder radius $R$.} (a) Case \ref{sec:5.1.3}; $\alpha =45.404023^\circ$. (b) Case \ref{sec:5.1.5}; \mbox{$\alpha = 45.40405^\circ$.} See text for details.}\label{fig:Max}
\end{figure}

\begin{figure}
  \centering
  \includegraphics[width=\columnwidth]{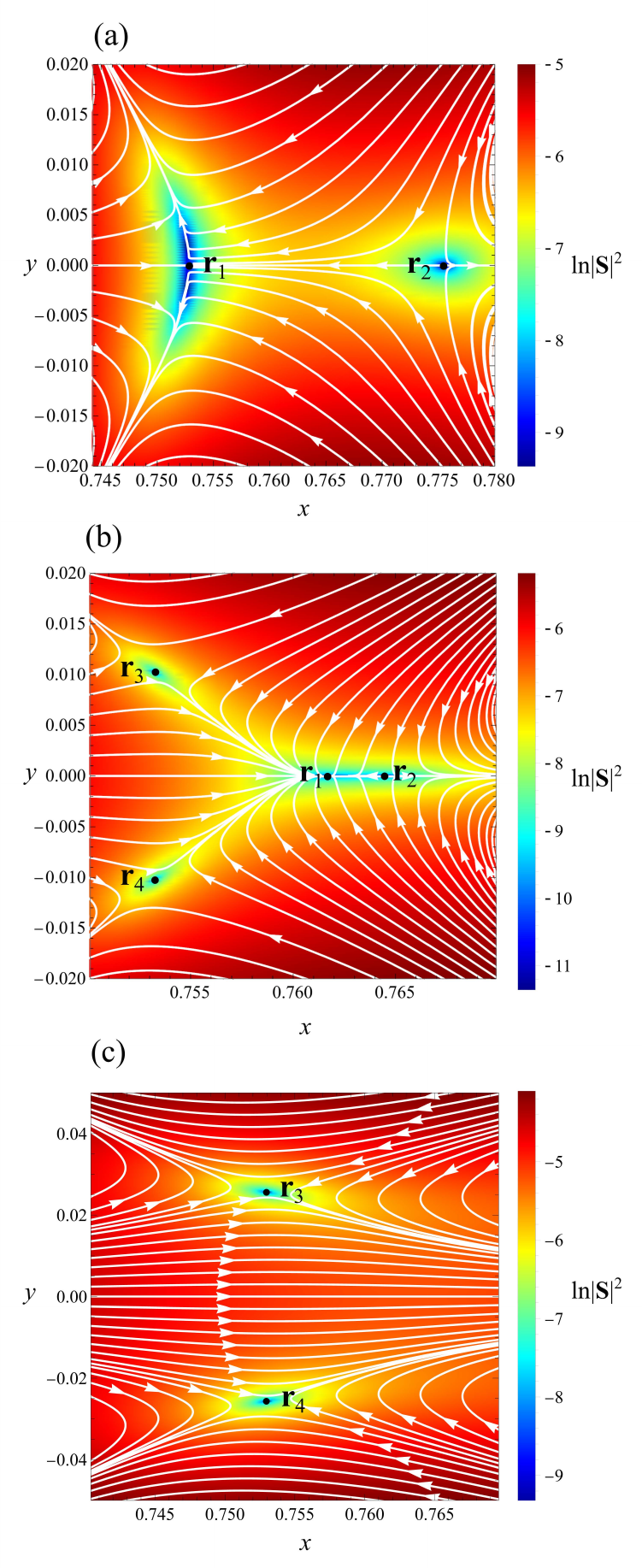}
  \caption{Projection of the streamlines onto the $xy$-plane. {The spatial coordinates are normalized on the cylinder radius $R$.} (a) Case \ref{sec:5.2.1}; \mbox{$kR =1.57311$.} (b) Case \ref{sec:5.2.3};  \mbox{$kR =1.573091$.} (c) Case \ref{sec:5.2.5}; $kR =1.573000$. See text for details.}\label{fig:Sad}
\end{figure}

Here we present the results of calculations, illustrating the above general discussion, in the case of a germanium right circular cylinder irradiated in a vacuum by a monochromatic linearly polarized plane electromagnetic wave with $\lambda=1590$~nm. According to Ref.~\cite{Polyanskiy}, the permittivity of germanium is taken as follows: \mbox{$\varepsilon = 17.775+0.024 i$.} In the vicinity of a local maximum of $S_x(x,y)$, we consider an oblique orientation of the polarization plane with respect to the cylinder axis. Thus, in this case, the control parameter is the angle $\alpha$; see Fig.~\ref{fig:Orientation}, while the radius of the cylinder's normal cross section is fixed so that the size parameter $kR=1.62$. Note that at this value of the size parameter $R/\lambda \approx 0.258$, i.e., the cylinder is a subwavelength scatterer.

{In contrast, to present the case of} a saddle point of $S_x(x,y)$, we consider the pure TM orientation of the incident wave, varying the radius of the cylinder's normal cross section so that, in this case, the size parameter is the bifurcation one.

{We must stress that the primary goal of the examples discussed below is to verify the developed phenomenological theory by comparing its predictions with the results of exact solutions of the scattering problem with actual values of permittivity for germanium. The mentioned values of the problem parameters are convenient for this goal since they make it possible to observe {\it all\/} discussed bifurcations, while the exact solutions (considered as a {\it virtual\/} experiment) easily allow to maintain any required accuracy of calculations. Then, the smallness of the differences between the obtained below bifurcation values of the control parameter, often lying beyond the actual experimental resolution, should not disappoint the reader.}

{Regarding the possibility of obtaining actual experimental evidence of the discussed phenomena, note that if the experimental accuracy is insufficient to resolve several sequential bifurcations individually, they are observed as a single merged bifurcation. Such a merged bifurcation also may be described by the developed phenomenology.} 

{Finally, note that the closeness of the bifurcation values of the control parameters in the discussed examples is not a generic feature of the problem. It is a peculiarity of the germanium cylinder at the specified values of $\lambda$ and $R$, whose choice was explained above. This is not the case for other materials, wavelengths, and radii. Then, if one wishes to obtain experimental evidence of a given bifurcation, the choice of the material, $\lambda$, and $R$, should be done accordingly.} 

{Let us proceed with discussing our ``virtual experiment'' (the exact solution to the problem at the specified values of the problem parameters). We have the following cases: }

%

\subsection{The vicinity of a maximum of $S_x(x,y)$\label{sec:6.1}} 

The oblique orientation of the polarization plane against the cylinder axis. The size parameter is fixed and equal to 1.62.
\begin{itemize}
  \item $\alpha = 45.403^\circ$. This value of $\alpha$ corresponds to case \ref{sec:5.1.1}. The projection of streamlines onto the $xy$-plane (the upper semiplane only) are shown in Fig.~\ref{fig:2D-3D}(a).
  \item $\alpha = 45.404023^\circ$. This case corresponds to Section \ref{sec:5.1.3}. The stable foci $\mathbf{r}_{3,4}$ have transformed into stable nodes; see Fig.~\ref{fig:Max}(a).
  \item $\alpha = 45.40405^\circ$. This is case \ref{sec:5.1.5}: points $\mathbf{r}_{3,4}$ have merged and annihilated, transferring their stable node type to point $\mathbf{r}_1$, while point $\mathbf{r}_2$ remains a saddle; see Fig.~\ref{fig:Max}(b).
\end{itemize}

\subsection{The vicinity of a saddle point of $S_x(x,y)$\label{sec:6.2}} 

The pure TM orientation of the incident wave. The size parameter varies.
\begin{itemize}
  \item $kR=1.57311$. This value of the size parameter corresponds to case \ref{sec:5.2.1}: only two singularities $\mathbf{r}_{1,2}$ exist. Both are saddles; see Fig.~\ref{fig:Sad}(a).
  \item $kR=1.573091$. At this value of $kR$ we have case \ref{sec:5.2.3}: four singular points; $\mathbf{r}_1$ is a stable node, $\mathbf{r}_{2,3,4}$ are saddles; see Fig.~\ref{fig:Sad}(b).
  \item $kR=1.573000$. Case \ref{sec:5.2.5}: only $\mathbf{r}_{3,4}$ remains. Both are saddles; see Fig.~\ref{fig:Sad}(c).
\end{itemize}

Regarding the dependence of various quantities on the corresponding control parameters discussed in this section, we have checked them in the proximity of both a maximum of $S_x$ and its saddle point. The calculations show that all predicted scalings  (see Sections \ref{sec:5.1.7}--\ref{sec:5.1.9}, \ref{sec:5.2.1}, and \ref{sec:5.2.3}, and the discussion in the text) hold within the employed numerical accuracy.
\vspace*{6pt}

\section{Conclusions\label{sec:conclus}}

The obtained results {have} revealed the deep connection between the topological structure of the Poynting vector field, the problem symmetry, and energy conservation law. Specifically, in the scattering of a monochromatic linearly polarized plane wave by an infinite right circular cylinder, a deviation of the orientation of the polarization plane from the pure TE or TM configuration makes the Poynting vector streamlines essentially three-dimensional. In this case, while the projection of the streamline pattern onto the plane of the cylinder's normal cross section may exhibit various types of singularities, in the three-dimensional space, all of them, but the ones {belonging to} the $x$-axis (see Fig.~\ref{fig:Orientation}), are regular points.

We have also shown that the bifurcation in the two-dimensional projection of streamline patterns may be explained and classified within the framework of the developed phenomenological theory. The latter is based on expanding the Poynting vector components in powers of the {deviations} of the coordinates from the specific characteristic points, supplemented by the constraints imposed by the symmetry and energy conservation law. All predictions made {by} this phenomenological theory have been illustrated and verified by analyzing the exact solution of the scattering problem for a germanium cylinder.

{We stress that the topological structure of the Poynting vector pattern changes only as a result of a finite number of bifurcations. It means that the discussed patterns are {\it topologically stable\/} at variations of the problem parameter(s) lying in between the bifurcation values. The conclusion should be valid for generalizing this discussion to any continuous symmetry violation in the light scattering problem. It ensures that the discussed topological properties of the problem are robust to a weak symmetry violation inevitable in any actual experiment.}    

The obtained results {unravel} essential features of the electromagnetic energy circulation at light scattering by particles (including nanoparticles). They shed new light on this problem and open the door to developing new nanotechnological approaches.
\mbox{}\\

\begin{acknowledgments}
The author is grateful to Boris Y. Rubinstein for valuable discussions and help in the visualization of the field patterns. This work was financially supported by the Russian Science Foundation under project no. 21-12-00151 (analytical research) and by the Ministry of Science and Higher Education of the Russian Federation under project no. 075-15-2022-1150 (computer calculations and computer graphics). The influence of symmetry effects on the properties of singular points of the Poynting vector was studied with the support of the Russian Science Foundation grant No. 23-72-00037.
\end{acknowledgments}

\bibliography{oblique_2} 

\end{document}